\documentclass[review]{elsarticle} 
\usepackage{hyperref}
\usepackage{comment} 
\usepackage{multirow}

\journal{Nuclear Instruments \& Methods in Physics Research, Section A}

\begin{document}
\begin{frontmatter}

\title{Picosecond timing of charged particles using the TORCH detector}

\author{M.~F.~Cicala$^1$}
\ead{maria.flavia.cicala@cern.ch}

\author{
S.~Bhasin$^{2,3}$,
T.~Blake$^1$,
N.~H.~Brook$^2$,
T.~Conneely$^4$,
D.~Cussans$^3$,
M.~W.~U.~van~Dijk$^5$,
R.~Forty$^5$,
C.~Frei$^5$,
E.~P.~M.~Gabriel$^6$,
R.~Gao$^7$,
T.~Gershon$^1$,
T.~Gys$^5$,
T.~Hadavizadeh$^7$,
T.~H.~Hancock$^7$,
N.~Harnew$^7$,
T.~Jones$^1$
M.~Kreps$^1$,
J.~Milnes$^4$, 
D.~Piedigrossi$^5$,
J.~Rademacker$^3$ and
J.~C.~Smallwood$^7$
}

\address{$^1$ Department of Physics, University of Warwick, Coventry, United Kingdom}
\address{$^2$ University of Bath, Claverton Down, Bath, United Kingdom}
\address{$^3$ H.H. Wills Physics Laboratory, University of Bristol, Bristol, United Kingdom}
\address{$^4$ Photek Ltd., St Leonards on Sea, East Sussex, United Kingdom}
\address{$^5$ European Organisation for Nuclear Research (CERN), Geneva, Switzerland}
\address{$^6$ School of Physics and Astronomy, University of Edinburgh, Edinburgh, United Kingdom}
\address{$^7$ Denys Wilkinson Laboratory, University of Oxford, Oxford, United Kingdom}

\begin{abstract}
TORCH is a large-area, high-precision time-of-flight (ToF) detector designed to provide charged-particle identification in the 2--20\,GeV/$c$ momentum range. Prompt Cherenkov photons emitted by charged hadrons as they traverse a 10\,mm quartz radiator are propagated to the periphery of the detector, where they are focused onto an array of micro-channel plate photomultiplier tubes (MCP-PMTs). The position and arrival times of the photons are used to infer the particles’ time of entry in the radiator, to identify hadrons based on their ToF. The MCP-PMTs were developed with an industrial partner to satisfy the stringent requirements of the TORCH detector. The requirements include a finely segmented anode, excellent time resolution, and a long lifetime. Over an approximately 10\,m flight distance, the difference in ToF between a kaon and a pion with 10\,GeV/$c$ momentum is $35$\,ps, leading to a 10--15\,ps per track timing resolution requirement. On average $30$ photons per hadron are detected, which translates to a single-photon time resolution of 70\,ps. 
The TORCH R$\&$D program aims to demonstrate the validity of the detector concept through laboratory and beam tests, results from which are presented. A timing resolution of 70-100\,ps was reached in beam tests, approaching the TORCH design goal. Laboratory timing tests consist of operating the MCP-PMTs coupled to the TORCH readout electronics. A time resolution of ~50\,ps was measured, meeting the TORCH target timing resolution.
\end{abstract}

\begin{keyword}
\texttt{Particle identification, Time-of-flight detectors} 
\end{keyword}

\end{frontmatter}

\section{Introduction}
\label{sec:introduction}

\begin{figure}[htb]
\begin{center}
\includegraphics[width=0.45\columnwidth]{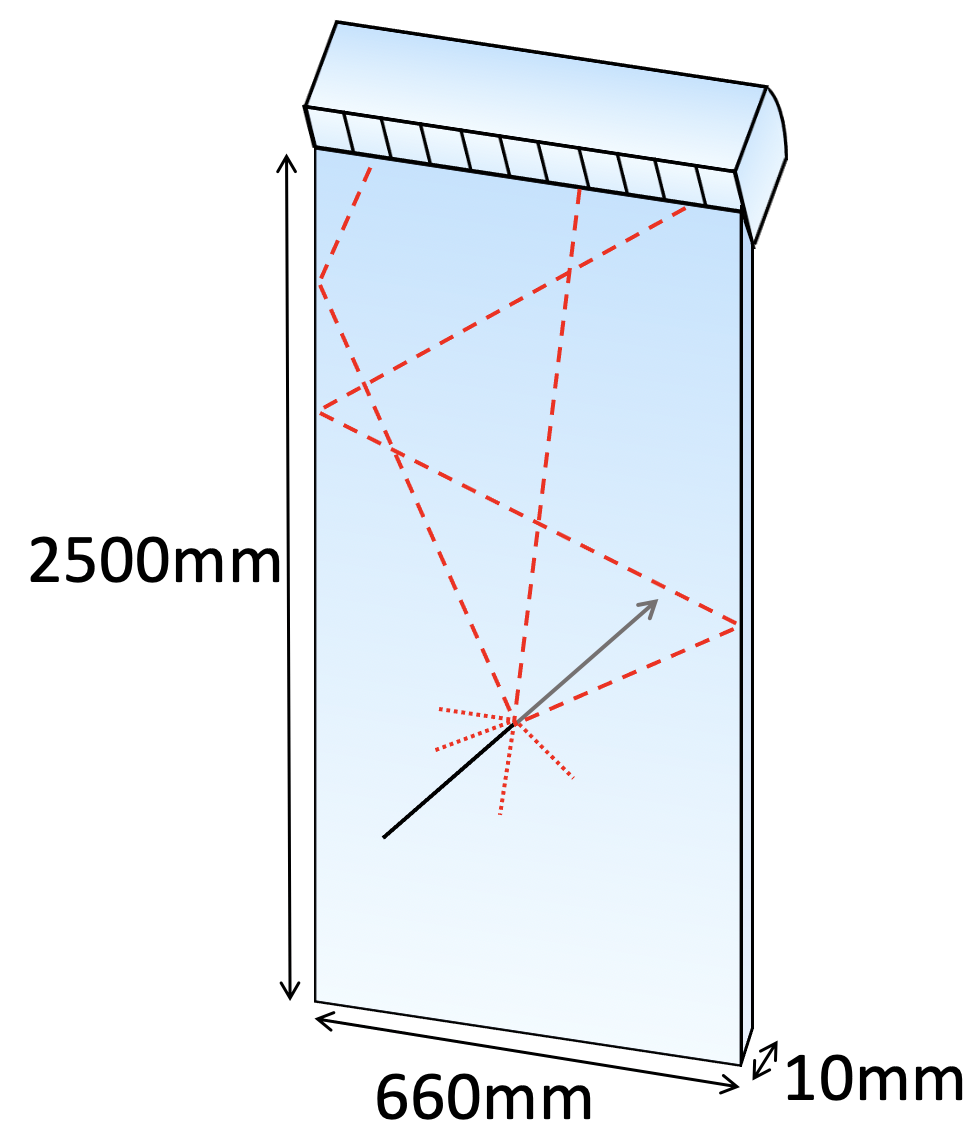} 
\includegraphics[width=0.45\columnwidth]{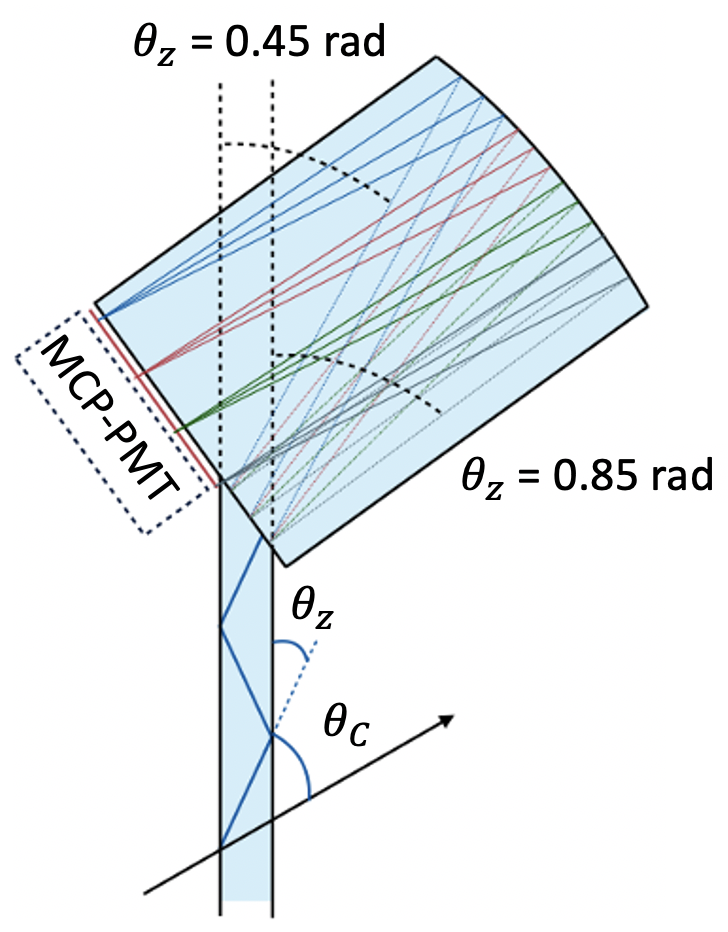}
\end{center} 
\caption{\label{Torch} Sketch illustrating a single TORCH module. Charged hadrons traversing the quartz plate emit Cherenkov photons, which are propagated to the edge of the detector via total internal reflection. The photons are focused by a cylindrical mirror onto a detector plane comprising 11 MCP-PMTs, when fully equipped. The position of a photon along the MCP-PMT is used to infer the photon Cherenkov angle.}
\end{figure}

The Time Of internally Reflected CHerenkov detector (TORCH) is a proposed time-of-flight (TOF) detector, which is designed to separate charged hadrons in the 2--20\,GeV/$c$ momentum range over a 10\,m flight path. The detector concept exploits the prompt Cherenkov radiation emitted when charged hadrons traverse a $10$\,mm thick quartz plate. Cherenkov photons are propagated to the periphery of the radiator, where they are reflected off a cylindrical-mirrored surface and are focused onto an array of micro-channel plate photomultiplier tubes (MCP-PMTs), as illustrated in figure~\ref{Torch}.

The time of entry of the hadron in the quartz can be calculated using information from a tracking system and the arrival time and position of the photons on the MCP-PMTs. The arrival position of the photons is used to calculate the Cherenkov angle and photon path length. The Cherenkov angle is used to correct for chromatic dispersion of photons in the quartz for the various particle hypotheses. From this and from the photons' arrival time, the time of propagation of the hadron and its time of flight is inferred~\cite{Charles:2010at}.

At a distance of $\sim 10$\,m from the production point, the difference in time-of-flight between a kaon and a pion with a momentum of 10\,GeV/$c$ is $\sim 35$\,ps. To statistically separate pions and kaons over this distance, a track timing resolution of 10--15\,ps is desirable. Hence, given an average of 30 detected photons per track, a single-photon time resolution of 70\,ps is required. The resolution receives contributions from the intrinsic time resolution of the MCP-PMT, the readout electronics, and the time resolution resulting from the photon path reconstruction. The target time resolution from the first contribution and from the combination of the second and third contributions is the same: around $50$\,ps per photon.

The TORCH programme involves an industrial partner, Photek (UK), for the development of the MCP-PMTs. The MCP-PMT prototypes are square tubes processed with atomic layer deposition (ALD), to tolerate an integrated charge accumulation $\geq 5$\,C/cm$^{2}$. The tubes have a finely segmented anode, providing the required granularity for the TORCH project. The MCP-PMTs have a $53 \times 53$\,mm$^2$ active area and are arranged linearly with a 60\,mm pitch to optimise the overall photon coverage. The anode segmentation is $64 \times 64$ pixels \cite{Conneely2015}.
Pixels are electronically ganged together in groups of eight in the non-focusing  (horizontal) direction, and charge sharing is exploited across the focusing direction (vertically) to achieve a $128$ × $8$ effective pixel granularity.

Connectivity between the anode output pads and readout electronics interface boards is achieved by the use of anisotropic conductive film.
The readout electronics~\cite{Gao2017} comprise the NINO \cite{Anghinolfi:2004gg} and HPTDC~\cite{Moreira2000} ASICs originally developed for the ALICE experiment~\cite{Despeisse5688208}. 
During data-taking, a single photoelectron is converted into a charged signal, shared amongst a cluster of pixels. The charge in each pixel is amplified and discriminated by the NINO. 
Information of the charge deposited in each pixel is measured by the time-over-threshold of the signal, which provides input to a cluster-algorithm. The HPTDC chip can be operated either in high resolution mode (100\,ps time bins) or very high resolution mode (25\,ps time bins). The signal is corrected for time-walk effects and for integrated non linearities of the HPTDC binning.

A version of TORCH detector has been proposed for PID at low momentum in the LHCb Upgrade II experiment \cite{LHCbCollaboration:2776420}. The simulated TORCH performance in the LHCb experiment in High Luminosity LHC (HL-LHC) conditions shows that TORCH can improve the overall LHCb detector particle identification (PID) performance, more details are given in section \ref{sec:PIDperformance}. 

The TORCH detector has been characterised in both laboratory and beam test environments. Two TORCH prototypes have been constructed, mini-TORCH \cite{Brook:2018qdc} and the larger scale proto-TORCH \cite{Smallwood:2021wed}, and have yielded encouraging results when exposed to low momentum charged hadrons in beam tests. Preliminary proto-TORCH time resolution results are detailed in section \ref{sec:testBeam}. In both beam tests, characteristic patterns of Cherenkov photons have been clearly recorded.

Complementary laboratory tests were carried out to investigate the gain, quantum efficiency, intrinsic time resolution, and time resolution coupled to the readout electronics of the MCP-PMTs. The latter tests are reported in section \ref{sec:labStudies}. All laboratory results demonstrate that the TORCH objectives are achievable.

\section{Beam test results}
\label{sec:testBeam}

The latest TORCH prototype (proto-TORCH) comprises a half height (125\,cm), full width (66\,cm) TORCH module, instrumented with two MCP-PMTs (Tube A and Tube B). The two tubes are located in the top left corner of the module when looking at its front face. Proto-TORCH underwent a beam test in October-November 2018 at the CERN PS East Hall T9 facility, with an 8\,GeV$/c$ beam composed of protons and pions. One of the purposes of the beam test was to measure the single-photon time resolution in proto-TORCH.
The beam-test setup comprised two threshold Cherenkov counters, two scintillator and time-reference stations, and a pixel telescope from EUDET/AIDA \cite{Rubinskiy2014}. The latter was used to measure the beam profile. The time reference stations comprise borosilicate bars from which Cherenkov light is detected using single channel MCP-PMTs. The stations were placed $10$\,m upstream and $1$\,m downstream of proto-TORCH. The threshold Cherenkov counters independently identify the particles in the beam. The HPTDC chips in proto-TORCH were operated in high resolution mode.
Single-photon time resolution measurements were performed, varying the beam incidence position on the quartz plate. Beam positions numbered 1, 3, 4 and 5 were $5$\,mm away from the edge of the plate closest to the MCP-PMTs and respectively $175$\,mm, $489$\,mm, $802$mm and $1115$mm distant from the top of the plate. Position 2 was $60$\,mm away from the edge and $175$\,mm from the top of the radiator. Position 6 was central along the horizontal axis and $1115$\,mm away from the top of the plate vertically. 

\begin{figure}[htb]
\begin{center}
\includegraphics[width=0.69\columnwidth]{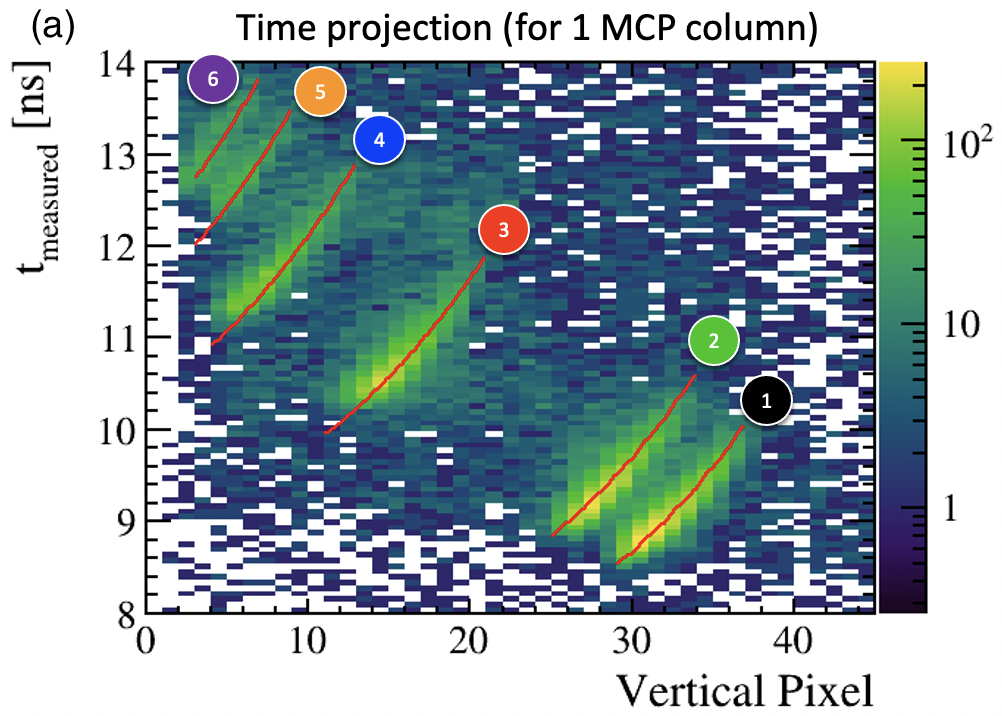} 
\includegraphics[width=0.30\columnwidth]{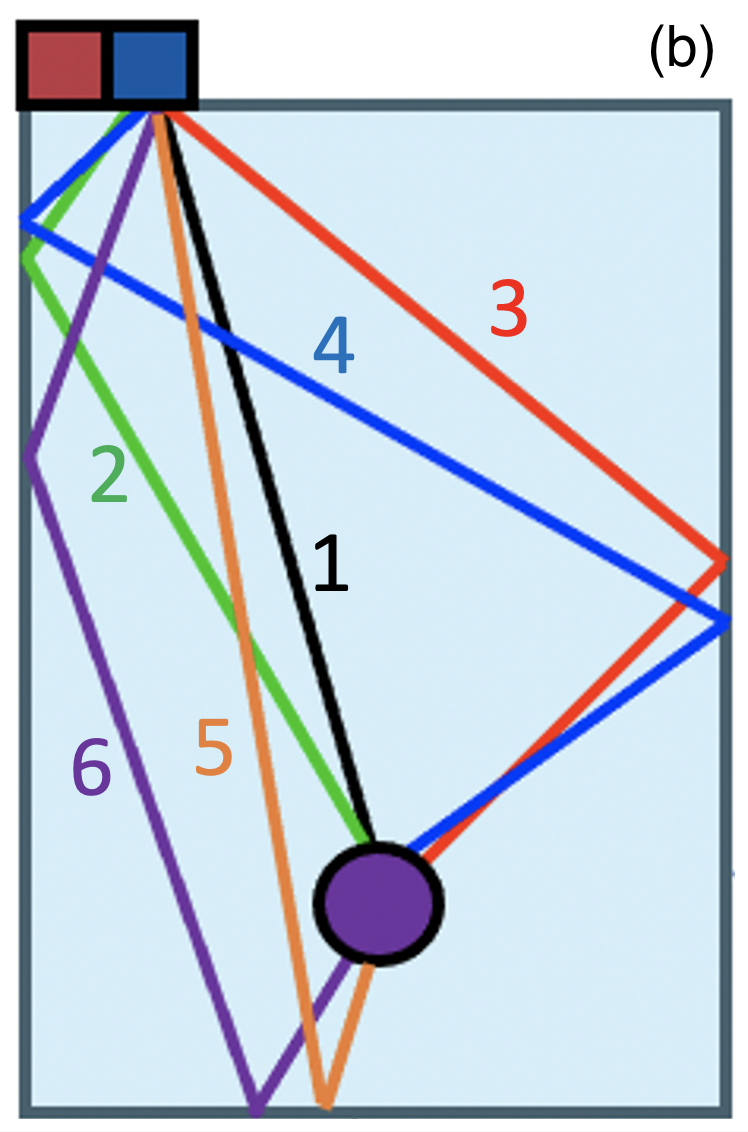}
\end{center} 
\caption{\label{TimeplotPaths} (a) Photon arrival time as a function of the vertical pixel number for a single horizontal pixel on MCP-PMT B. Six distinct bands are visible, corresponding to the six paths illustrated in (b). The bands are overlaid with the expected arrival postion/time from simulation. The width of the bands in data corresponds to the single-photon time resolution.}
\end{figure}

\begin{figure}[htb]
\begin{center}
\includegraphics[width=0.63\columnwidth]{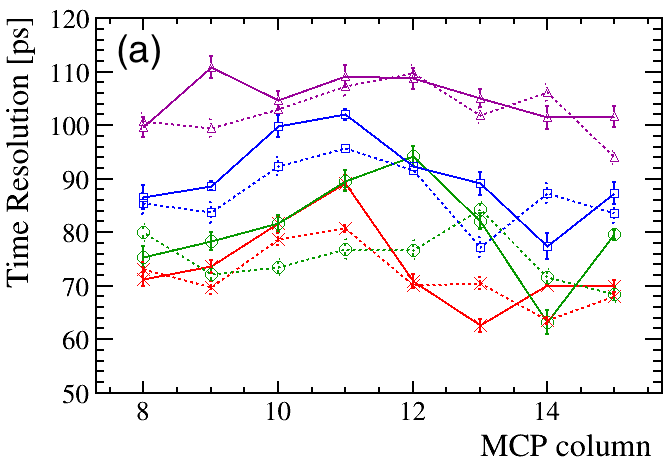} 
\caption{\label{SingleTR2D} 
TORCH single-photon time resolution, as a function of MCP B column number, measured for pions (continuos line)  and protons (dashed line) for a particular photon path, for different beam entry positions: 1 (red  $\times$), 3 (green $\circ$), 4 (blue \fbox{\phantom{\rule{.2ex}{.2ex}}} ) and 5 (purple $\triangle$).
}
\end{center}
\end{figure}

Photons in a direct path to the MCP-PMTs, reflection-less from the radiator sides, land on the detection plane first, mapping part of the Cherenkov cone visible in the raw hitmap. The remaining part of the Cherenkov cone is folded by the side-reflections, creating a criss-cross pattern on the raw hitmap. Photons which are reflected off the sides of the quartz bar have longer paths and arrive later in time. Different paths are visible in figure \ref{TimeplotPaths} (b). These are labelled 1-6 and should not be confused with the beam positions. The overall effect is that photons that experience the same number of side-reflections are detected at a similar time, with a longer time interval between the arrival of photons experiencing a different number of side-reflections.
The collected data can be rearranged according to the arrival time of the photons. This is illustrated in figure \ref{TimeplotPaths} where, for a single horizontal pixel, distinct bands are visible for a column of pixels of MCP-PMT B. A single-photon time resolution measurement is made by measuring the width of the band in the vertical pixel and photon arrival time space. For each reflection combination, the distribution of the photon arrival time in each vertical pixel is fitted with a probability density function (PDF) to extract the measured time resolution, $\sigma_{\rm measured}$. The PDF comprises a polynomial function to model the background and Crystal Ball function to model the peaking structures. The measured resolution is corrected for the time resolution of the time reference stations, $\sigma_{\rm ref.}$, and for the finite size of the beam through the quartz plate, $\sigma_{\rm beam}$, to calculate the resolution of the TORCH system:
\begin{displaymath}
     \sigma^2_{\rm TORCH} = \sigma^2_{\rm measured} - \sigma^2_{\rm ref.} - \sigma^2_{\rm beam}.
\end{displaymath}
The resulting single-photon time resolution for different beam incidence positions, for a particular photon path, is shown in figure \ref{SingleTR2D}. The resolution approaches the design goal of $70$\,ps for beam positions closer to the detector plane. The time resolution is degraded for beam incidence positions which are farther from the detector plane due to chromatic dispersion and the longer photon path length.

The TORCH single-photon time resolution can be factorised into three components~\cite{Hadavizadeh:2019pby} 
\begin{displaymath}
     \sigma^2_{\rm TORCH} = \sigma_{\rm prop}^2 (t_{\rm p}) + \sigma_{\rm RO}^2 (\sqrt{N_{\rm hits}}) + \sigma^2_{\rm MCP-PMT}~,
\end{displaymath}
where $\sigma_{\rm prop} (t_{\rm prop})$ is the time resolution on the measurement of the photon propagation time in the quartz, $t_{\rm prop}$, where $\sigma_{\rm MCP-PMT}$ is the MCP-PMT's intrinsic time resolution, and $\sigma_{\rm RO} (\sqrt{N_{\rm hits}})$ is time resolution associated with the readout electronics. The latter scales by $\sqrt{N_{\rm hits}}$, the number of pixels hit by a single photon. The time resolution component related to the photon propagation time in the quartz is expected to grow linearly with the propagation time due to the fixed pixel size, and $\sim 50$\,ps is targeted. The contribution of the MCP-PMT intrinsic time resolution and the readout electronics time resolution is expected to be $\sim 50$\,ps. This is investigated in section~\ref{sec:labStudies}. The measured values from the proto-TORCH beam test for the three TORCH single-photon time resolution components are shown in table~\ref{tab:ttr2d}.

\begin{table}[htb]
    \centering
    \begin{tabular}{ l | l | l }
        \hline
        Contribution & Measured [ps] & Simulation [ps] \\
        \hline
        $\sigma_{\rm prop} (t_{\rm prop})$ & $(7.6\pm 0.5)\times t_p [\rm ns]$ & $(3.8\pm 0.8)\times t_p [\rm ns]$\\
        $\sigma_{\rm MCP-PMT}$ & $31.0\pm7.6$ & 33\\
        $\sigma_{\rm RO}(\sqrt{N_{\rm hits}})$ & $\frac{95\pm 6}{(\sqrt{N_{\rm hits}})} $ & $\frac{60}{(\sqrt{N_{\rm hits}})}$\\
        \hline
    \end{tabular}
    \caption{Comparison between measured and predicted contributions to the TORCH time resolution. The components are described in the text.}
    \label{tab:ttr2d}
\end{table}

\section{Laboratory studies of time resolution}
\label{sec:labStudies}

\begin{figure}[htb]
\begin{center}
\includegraphics[width=0.60\columnwidth]{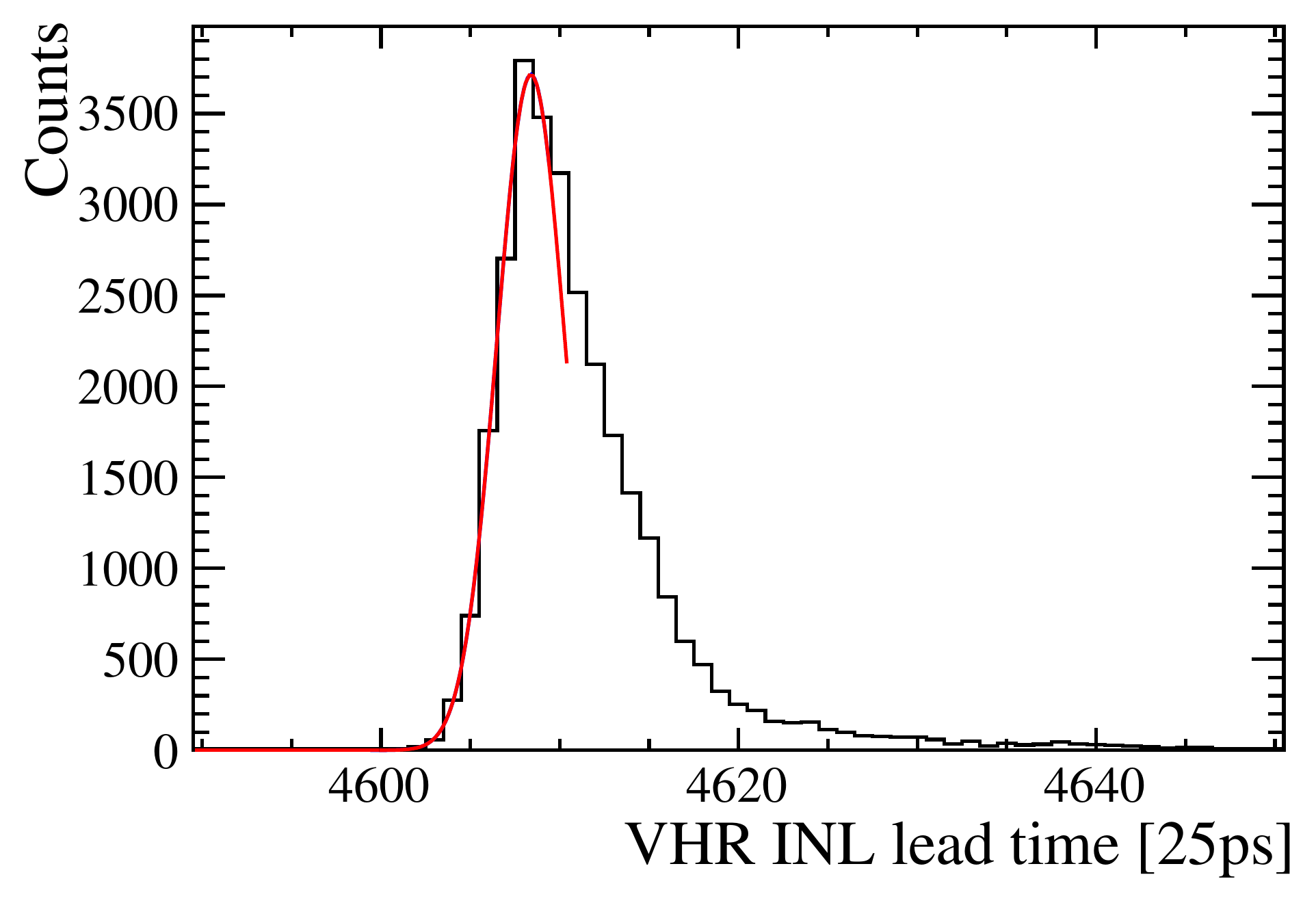}
\caption{\label{VHR_INLcorrected_constantAmplitude} 
Leading edge distribution of constant time-over-threshold signals from the TORCH readout chain, selected from 10 million events collected with the laboratory setup described in the text. The left-hand side of the distribution is fitted with a Gaussian function. 
}
\end{center}
\end{figure}

A laboratory test-stand was set up to measure the combined time resolution of the MCP-PMT and the TORCH readout electronics. The MCP-PMT was placed in a light-tight box and illuminated with a fast-pulsed blue laser source. The test stand can be operated in two configurations: the laser light can be focused as a spot with a diameter of $\sim 20\, {\rm \mu m}$; or the laser light can be diffused to illuminate the entire tube for HPTDC calibration studies and for homogeneity studies. The trigger of the readout electronics was synchronised with the laser pulses. The HPTDC chip was operated in the very high resolution mode and the MCP-PMT operated at a gain of  $\mathcal{O}(10^6)$. 
The integrated non-linear response of the HPTDC chip was corrected by using calibration data with the diffused light source. Triggered events with constant time-over-threshold values are assumed to have constant amplitude to first order, and were selected to eliminate time-walk effects. The leading edge distribution of the events was fitted with a Gaussian function, as illustrated in figure \ref{VHR_INLcorrected_constantAmplitude}, to estimate the time resolution. The function has a standard deviation of $(47.5 \pm 0.7)$\,ps, compatible with the 50\,ps TORCH goal. The fit range does not cover the tail of the distribution as it is skewed by the relaxation pulse of the laser and by electron back-scattering from the MCP-PMT input face. The same measurement was performed without the requirement of constant time-over-threshold values and the resulting time resolution is $(63.2 \pm 0.2)$\,ps.

\section{PID Performance}
\label{sec:PIDperformance}
\begin{figure}[htb]
\begin{center}
\includegraphics[width=0.70\columnwidth]{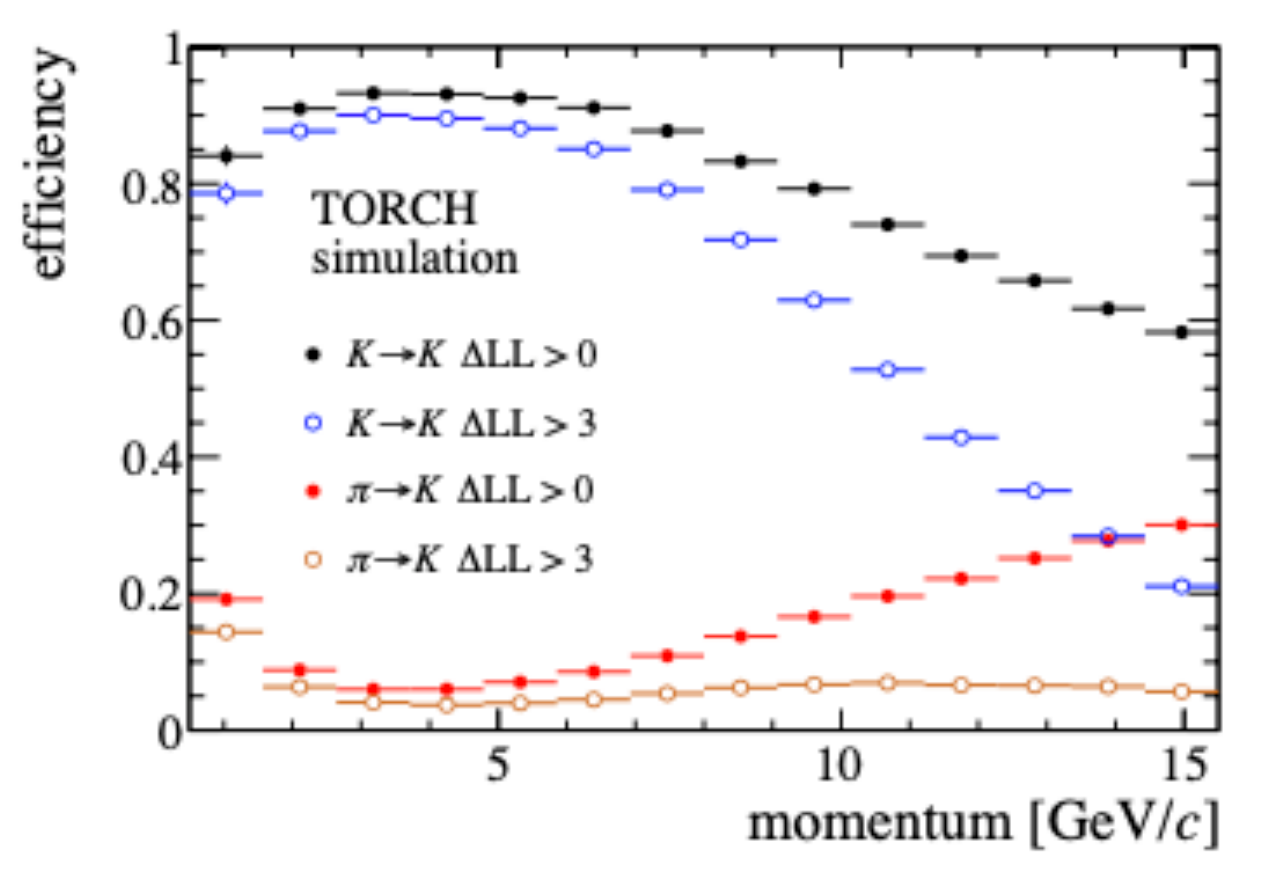} 
\includegraphics[width=0.70\columnwidth]{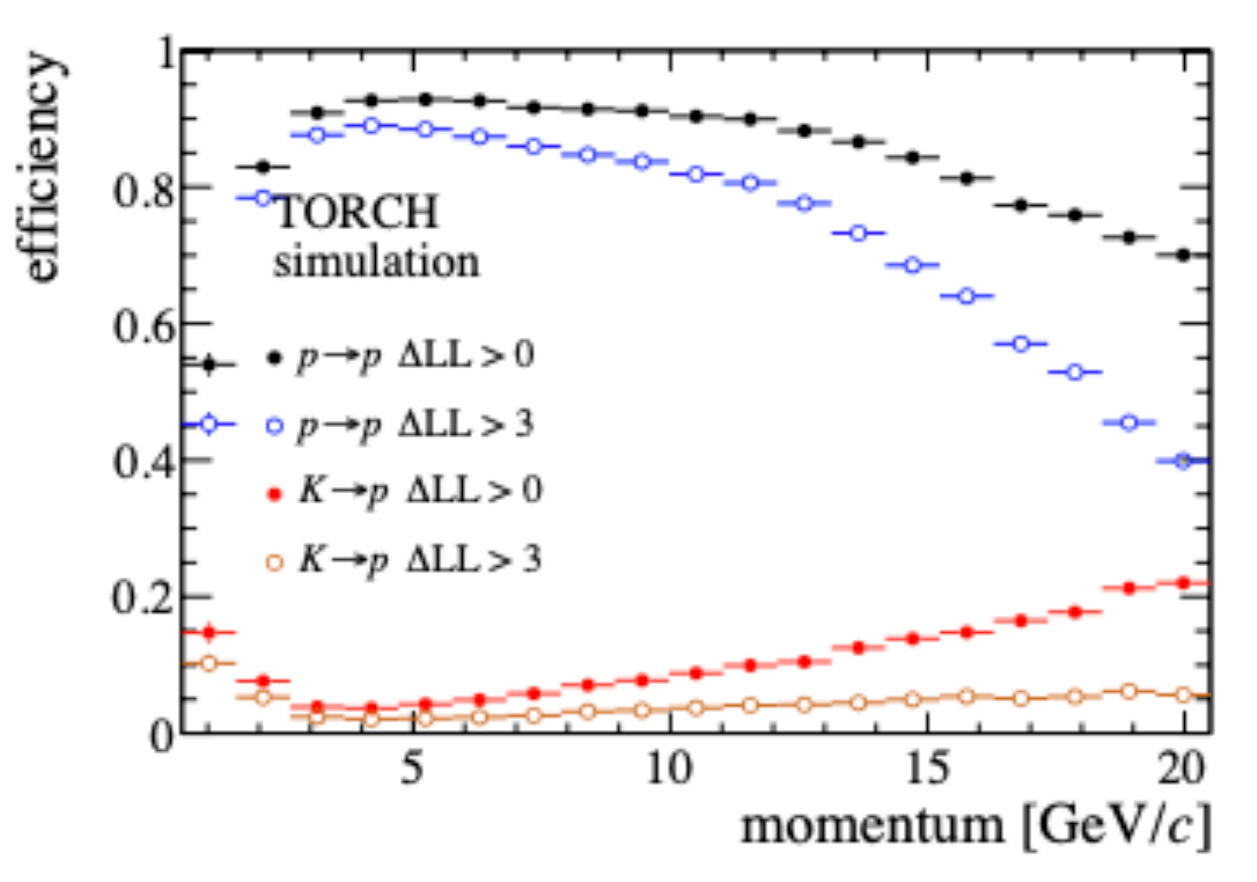}
\caption{\label{PID2} TORCH PID performance in LHCb in HL-LHC data taking conditions. Two different likelihood requirements are shown for kaon-pion separation efficiency (top) and for proton-kaon separation efficiency (bottom).}
\end{center}
\end{figure}

A TORCH detector composed of 18 modules has been proposed for PID at low momentum in the LHCb Upgrade II experiment \cite{LHCbCollaboration:2776420}. It would be positioned at a $\sim 9.5$\,m distance from the proton-proton interaction point at LHCb. The predicted PID performance of the TORCH detector as a part of the LHCb experiment was assessed by simulating the 18 TORCH modules in the LHCb GEANT4 simulation. The TORCH simulation software is modelled on the LHCb's RICH detector reconstruction software, where PID for a given track is achieved by comparing the likelihood of each hadron species hypothesis with the photon pattern caused by the track \cite{Forty:1998eqa}. 
High-luminosity LHC conditions were simulated. These correspond to $\mathcal{L} =  1.4 \times 10^{34}\, \mbox{cm}^{-2} \mbox{s}^{-1}$ and to a TORCH MCP-PMT effective pixel granularity of $128 \times 32$ and of $128 \times 8$, for the central and peripheral modules, respectively. \cite{LHCbCollaboration:2776420}. Figure \ref{PID2} illustrates the TORCH kaon-pion (proton-kaon) separation power achievable.

\section{Conclusions}
\label{sec:conclusions}

TORCH is a large scale time-of-flight detector aiming to provide charged particle separation in the 2--20\,GeV/$c$ momentum range over a 10\,m flight path. A full-width, half-height, partially instrumented TORCH prototype module underwent a successful test beam in 2018; a single-photon time resolution that approaches the design goal of $70$\,ps required to obtain a $10-15$\,ps per-track timing resolution, was measured. Laboratory studies of the time resolution of the TORCH MCP-PMT and readout electronics complement the test beam results. A measured resolution of  $(47.5 \pm 0.7)$\,ps shows that TORCH time resolution design goals are achievable. Simulation studies of the TORCH detector in the LHCb experiment show that TORCH can significantly add to the LHCb physics program by providing efficient pion-kaon and kaon-proton separation in the 2--10\,GeV$/c$ and 2--20\,GeV$/c$ momentum ranges, respectively.

The test beam and laboratory studies of the TORCH time resolution demonstrate good performance which is approaching the TORCH design goals. Better calibration methods are currently under study and a future test beam campaign is planned for 2022 with the fully equipped TORCH prototype module.

\section*{Acknowledgements}

The support is acknowledged of the Science and Technology Research Council, UK, grant number ST/P002692/1, of the European Research Council through an FP7 Advanced Grant (ERC-2011-AdG 299175-TORCH) and of the Royal Society, UK. 

\bibliography{references}

\begin{thebibliography}{10}
\expandafter\ifx\csname url\endcsname\relax
  \def\url#1{\texttt{#1}}\fi
\expandafter\ifx\csname urlprefix\endcsname\relax\def\urlprefix{URL }\fi
\expandafter\ifx\csname href\endcsname\relax
  \def\href#1#2{#2} \def\path#1{#1}\fi

\bibitem{Charles:2010at}
M.~J. Charles, R.~Forty, {TORCH: Time of Flight Identification with Cherenkov
  Radiation}, Nucl. Instrum. Meth. A 639 (2011) 173--176.
\newblock \href {https://doi.org/10.1016/j.nima.2010.09.021}
  {\path{doi:10.1016/j.nima.2010.09.021}}.

\bibitem{Conneely2015}
T.~M. Conneely, et~al., {The TORCH PMT: A close packing, multi-anode, long life
  MCP-PMT for Cherenkov applications}, JINST 10~(5) (2015) C05003.
\newblock \href {https://doi.org/10.1088/1748-0221/10/05/C05003}
  {\path{doi:10.1088/1748-0221/10/05/C05003}}.

\bibitem{Gao2017}
R.~Gao, et~al., Precision electronics for a system of custom mcps in the torch
  time of flight detector, Journal of Instrumentation 12 (2017) C03008--C03008.
\newblock \href {https://doi.org/10.1088/1748-0221/12/03/C03008}
  {\path{doi:10.1088/1748-0221/12/03/C03008}}.

\bibitem{Anghinolfi:2004gg}
F.~Anghinolfi, P.~Jarron, A.~N. Martemyanov, E.~Usenko, H.~Wenninger, M.~C.~S.
  Williams, A.~Zichichi, {NINO: An ultra-fast and low-power front-end
  amplifier/discriminator ASIC designed for the multigap resistive plate
  chamber}, Nucl. Instrum. Meth. A 533 (2004) 183--187.
\newblock \href {https://doi.org/10.1016/j.nima.2004.07.024}
  {\path{doi:10.1016/j.nima.2004.07.024}}.

\bibitem{Moreira2000}
M.~Mota, et~al., {A flexible multi-channel high- resolution time-to-digital
  converter ASIC}, IEEE Nuclear Science Symp. Conf. Record 2 (2000) 155--9.

\bibitem{Despeisse5688208}
M.~Despeisse, F.~Powolny, P.~Jarron, J.~Lapington, Multi-channel
  amplifier-discriminator for highly time-resolved detection, IEEE Transactions
  on Nuclear Science 58~(1) (2011) 202--208.
\newblock \href {https://doi.org/10.1109/TNS.2010.2100409}
  {\path{doi:10.1109/TNS.2010.2100409}}.

\bibitem{LHCbCollaboration:2776420}
{LHCb Collaboration}, {Framework TDR for the LHCb Upgrade II - Opportunities in
  flavour physics, and beyond, in the HL-LHC era },
  \url{http://cds.cern.ch/record/2776420} (Jul 2021).

\bibitem{Brook:2018qdc}
N.~Brook, et~al., {Testbeam studies of a TORCH prototype detector}, Nucl.
  Instrum. Meth. A 908 (2018) 256--268.
\newblock \href {https://doi.org/10.1016/j.nima.2018.07.023}
  {\path{doi:10.1016/j.nima.2018.07.023}}.

\bibitem{Smallwood:2021wed}
J.~C. Smallwood, et~al., \href{http://cds.cern.ch/record/2790152/}{{Test-beam
  demonstration of a TORCH prototype module}}, in: {International Conference on
  Technology and Instrumentation in Particle Physics}, 2021.
\newline\urlprefix\url{http://cds.cern.ch/record/2790152/}

\bibitem{Rubinskiy2014}
I.~Rubinskiy, H.~Perrey, An eudet/aida pixel beam telescope for detector
  development, Proc. of Science 213 (2014) 122.

\bibitem{Hadavizadeh:2019pby}
T.~Hadavizadeh, et~al., {Status of the TORCH time-of-flight detector}, PoS
  EPS-HEP2019 (2020) 140.
\newblock \href {https://doi.org/10.22323/1.364.0140}
  {\path{doi:10.22323/1.364.0140}}.

\bibitem{Forty:1998eqa}
R.~W. Forty, O.~Schneider, \href{http://cds.cern.ch/record/684714}{{RICH
  pattern recognition}} (4 1998).
\newline\urlprefix\url{http://cds.cern.ch/record/684714}

\end{thebibliography}
\end{document}